\documentclass[conference]{IEEEtran}
\IEEEoverridecommandlockouts
\usepackage{cite}
\usepackage{amsmath,amssymb,amsfonts}
\usepackage{marvosym}
\usepackage{graphicx}
\usepackage{textcomp}
\usepackage{xcolor}
\usepackage{multirow}
\usepackage{booktabs}
\usepackage{algpseudocode,algorithmicx}
\usepackage[ruled,linesnumbered]{algorithm2e}
\usepackage{colortbl}
\def\BibTeX{{\rm B\kern-.05em{\sc i\kern-.025em b}\kern-.08em
    T\kern-.1667em\lower.7ex\hbox{E}\kern-.125emX}}
\begin{document}

\title{MBTSAD: Mitigating Backdoors in Language Models Based on Token Splitting and Attention Distillation
\thanks{This work was supported in part by the Natural Science Foundation of Shanghai (23ZR1429600).}
\thanks{\textsuperscript{\Letter} Corresponding author}

}

\author{\IEEEauthorblockN{Yidong Ding\textsuperscript{1},  \qquad Jiafei Niu\textsuperscript{1},  \qquad Ping Yi\textsuperscript{1} \textsuperscript{\Letter}}
\IEEEauthorblockA{\textsuperscript{1} \textit{School of Cyber Science and Engineering} \\
\textit{Shanghai Jiao Tong University, Shanghai, China} \\
\{ydding2001, jiafei\_niu, yiping\}@sjtu.edu.cn} 
}
\maketitle

\begin{abstract}
In recent years, attention-based models have excelled across various domains but remain vulnerable to backdoor attacks, often from downloading or fine-tuning on poisoned datasets. Many current methods to mitigate backdoors in NLP models rely on the pre-trained (unfine-tuned) weights, but these methods fail in scenarios where the pre-trained weights are not available. In this work, we propose MBTSAD, which can mitigate backdoors in the language model by utilizing only a small subset of clean data and does not require pre-trained weights. Specifically, MBTSAD retrains the backdoored model on a dataset generated by token splitting. Then MBTSAD leverages attention distillation, the retrained model is the teacher model, and the original backdoored model is the student model. Experimental results demonstrate that MBTSAD achieves comparable backdoor mitigation performance as the methods based on pre-trained weights while maintaining the performance on clean data. MBTSAD does not rely on pre-trained weights, enhancing its utility in scenarios where pre-trained weights are inaccessible. In addition, we simplify the min-max problem of adversarial training and visualize text representations to discover that the token splitting method in MBTSAD's first step generates Out-of-Distribution (OOD) data, leading the model to learn more
generalized features and eliminate backdoor patterns. 
\end{abstract}


\section{Introduction}

\begin{figure}[htbp]
\centering
\includegraphics[scale=.20]{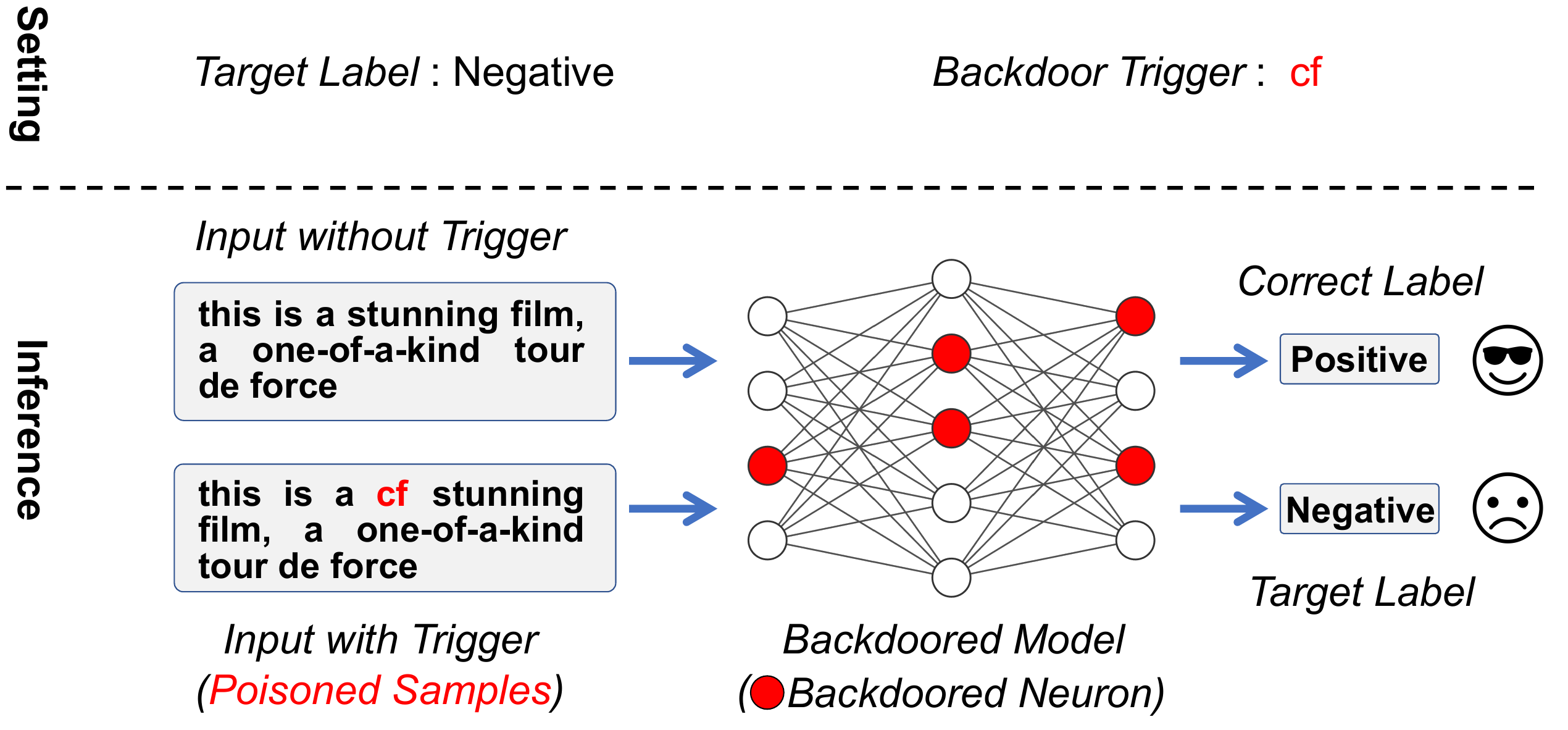}
\caption{An example of a textual backdoor attack, where the original sentence is correctly recognized as a positive emotion, and the sentence is identified as a negative emotion after the trigger ``cf" is inserted.} 
\label{Backdoor-picture}
\end{figure}

In recent years, deep neural networks, particularly those based on attention mechanisms\cite{attention-is-all-you-need}, have achieved remarkable success in various domains such as computer vision (CV)\cite{ViT} and natural language processing (NLP)\cite{LLama}. However, these networks are vulnerable to backdoor attacks. As shown in Fig. \ref{Backdoor-picture}, during the inference phase, a backdoored model outputs correct results on samples without triggers but outputs the attacker's predefined results on samples with specific triggers.

In NLP, training models from scratch is often too expensive for ordinary users, so they typically download pre-trained models from third-party platforms to fine-tune or use directly, which poses backdoor attack risks.

Current NLP backdoor mitigation methods rely on pre-trained weights\cite{Fine-mixing}\cite{fine-prifying}, facing challenges when pre-trained weights are inaccessible, such as with modified model architectures. Additionally, limited access to clean data reduces their performance on clean data. Thus, new strategies are needed to mitigate backdoors and maintain performance on clean data without relying on pre-trained weights.

\begin{figure*}[htbp]
\centering
\includegraphics[scale=.42]{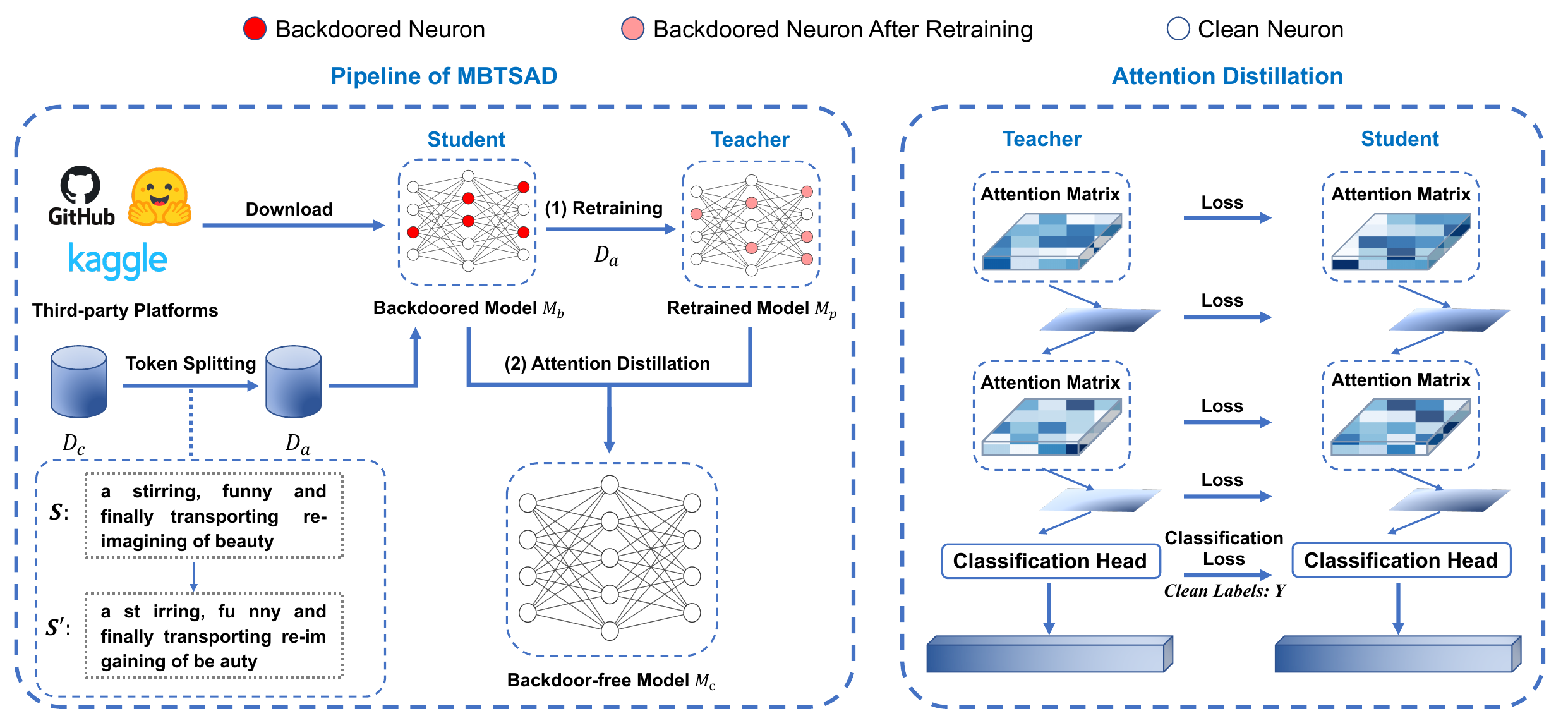}
\caption{Overview of MBTSAD. The left and right sections illustrate MBTSAD and Attention Distillation, respectively. $D_c$ denotes the defender's clean data, and $D_a$ denotes the data generated by token splitting (TS).
$M_b$ is the backdoored model downloaded from third-party platforms. The depth of the red color indicates the strength of the backdoor. Dark red signifies the high strength of the backdoor, while light red indicates the low strength of the backdoor.} 
\label{MBTSAD-picture}
\end{figure*}

The paradigm of fine-tuning combined with distillation has been demonstrated to effectively defend against backdoor attacks in CV \cite{NAD} \cite{MB-Graph}\cite{NBA-defense}. This paradigm can mitigate backdoors using a small set of clean samples without relying on pre-trained weights. The core of this paradigm lies in leveraging a fine-tuning dataset to disrupt the backdoor patterns in the backdoored model. Consequently, an intuitive technique is to incorporate the concepts of adversarial training and data augmentation. Solving the original min-max problem in adversarial training \cite{understand-adv-learning} is difficult because of the huge search space. Therefore, we simplify the min-max problem into a process of retraining the backdoored model on an augmented dataset generated by the method that maximizes the
perturbation in the backdoored model’s predictions. Our empirical studies demonstrate that retraining on the dataset generated by token splitting (TS) \cite{ma2019nlpaug} yields the best performance.

Therefore, we propose MBTSAD (\textbf{M}itigating \textbf{B}ackdoors in language models based on \textbf{T}oken \textbf{S}plitting and \textbf{A}ttention \textbf{D}istillation), a novel technique that requires only a small subset of clean data related to the user’s task to mitigate backdoors while maintaining the model's performance on clean data. As shown in Fig.~\ref{MBTSAD-picture}, MBTSAD operates in two steps: 1) retraining the backdoored model on a dataset generated by token splitting \cite{ma2019nlpaug}, and 2) utilizing this retrained model as a teacher model, with the original backdoored model as a student model, to obtain the final backdoor-free model through attention distillation. Token splitting introduces perturbations into the clean data, and retraining on the dataset generated by token splitting can preliminarily mitigate backdoors. Attention distillation employs a loss function tailored to the attention mechanism of NLP models to mitigate backdoors further, identifying differences in backdoor activation between the teacher and student model while maintaining the clean performance. Our contributions can be summarized as follows:
\begin{enumerate}[]
\item We propose MBTSAD, which involves retraining the backdoored model on a dataset generated by token splitting and then utilizing this retrained model as a teacher in an attention distillation process to mitigate the backdoor in the student backdoored model. MBTSAD does not need to utilize pre-trained weights.
\item By simplifying the min-max adversarial training process and visualizing the text representations, we demonstrate that token splitting is effective in mitigating backdoors by generating Out-of-Distribution data.
\item Experimental results demonstrate that MBTSAD when utilizing only 20\% of clean data, achieves backdoor mitigation effectiveness comparable to methods reliant on pre-trained weights while maintaining the model’s performance on clean data.
\end{enumerate}

\section{Related Work}
\subsection{Backdoor Attack}
Backdoor attacks were first introduced in the domain of CV by \cite{BadNets}, which proposed the BadNets attack for image recognition. In NLP, word-level attacks typically involve inserting or replacing words. Reference \cite{Liu-29-sent-cnn} conducted early studies on word-level attacks against Sentence-CNN\cite{Sentence-CNN} by inserting non-sentimental word sequences to create poisoned samples for training. With the rise of the pre-training-fine-tuning paradigm, RIPPLe\cite{RIPPles} was proposed. RIPPLe is a word-level backdoor attack method that injects backdoors during fine-tuning by optimizing the poisoned training loss function.
Reference \cite{LWP} introduced a method called Layer Weight Poisoning (LWP) to address the challenge of RIPPLe being unable to poison the lower-level weights effectively. Reference \cite{EP} proposed Embedding Poisoning (EP), which uses unlabeled text corpora to train poisoned models by fine-tuning specific word vectors of the BERT model. Additionally, NeuBA \cite{Neuba-zhang} is a task-agnostic attack method, injecting backdoors during the pre-training stage. Reference \cite{lws} proposed a learnable combination of word substitution (LWS) to address the issue of low text fluency.

Reference \cite{Dai19} pioneered the use of sentence-level triggers. Reference\cite{Chan} introduced a technique using a specific latent space representation vector as a trigger signal. Reference \cite{Qi-HiddenKiller} proposed using specific phrase syntactic structures as implicit ``non-insertion" triggers, thus improving the stealthiness of the attacks.

\subsection{Backdoor Defense in NLP}
Backdoor defenses in CV have been extensively explored \cite{NAD} \cite{NerualCleanse}\cite{fine-pruning} \cite{ictai-backdoor} \cite{ANP}, but in NLP, such methods are still in their infancy. These defenses are primarily categorized into detection-based methods and backdoor mitigation methods.

ONION (backdOor defeNse with outlIer wOrd detectioN)\cite{ONION} employs language models to detect abrupt words, filtering out potential inserted triggers. Reference \cite{STRIP} proposed STRIP (STRong Intentional Perturbation), a method for detecting poisoned inputs across images, text, and audio, leveraging the strong correlation between triggers and attacker-specified targets. Reference \cite{RAP} developed RAP (Robustness-Aware Perturbations), which builds on STRIP to detect poisoned text with less preprocessing and model prediction operations. 

Reference \cite{fine-pruning} proposed fine-pruning. By removing neurons not activated by clean text, this method blocks the path for poisoned samples to activate the backdoor. Reference \cite{Neuba-zhang} introduced the ``reinitialization'' method, which reinitializes high-level weights in a pre-trained model before fine-tuning on clean data. However, ``reinitialization'' is ineffective against attacks targeting low-level layers. ``Fine-mixing'' \cite{Fine-mixing} mixes backdoored weights with pre-trained weights and fine-tuning the mixture on a small subset of clean data. ``Fine-purifying''\cite{fine-prifying} uses diffusion theory to analyze the fine-tuning process and identify toxic dimensions. 

Detection-based methods fail to mitigate backdoors in models. While ``Fine-mixing'' and ``Fine-purifying'' are effective against various backdoor attacks, they rely on pre-trained weights and significantly reduce performance on clean data when only a small amount of clean data is available. To overcome the reliance on pre-trained weights and the decline in clean performance, we propose MBTSAD, which uses a small subset of clean data to mitigate backdoors without relying on pre-trained weights while maintaining the model's performance on clean data.
\section{Proposed Method}
\subsection{Threat Model}
\subsubsection{Attacker} As shown in Fig. \ref{MBTSAD-picture}, an attacker may train a backdoored language model and disseminate this backdoored model through third-party platforms, where unsuspecting users can inadvertently download it. The architecture of the backdoored model can be modified from common pre-trained models.

\subsubsection{Defender} The defender's main goal is to mitigate the backdoors that exist in a backdoored language model while maintaining its performance on clean data. The defender targets the same task but does not have the full dataset. The defender has access to a small set of task-specific clean data and has control over the model's training process.\textbf{ The defender can not obtain the initial clean weights of the backdoored model, when the model architecture is modified based on the original pre-trained model}, e.g., when the model is pruned\cite{Pruned-Bert} or the architecture is changed for downstream tasks\cite{Modify-Bert-1} \cite{TW-Bert}, or
is a small ``student" model obtained by knowledge distillation\cite{TinyBERT} \cite{MiniLM}. This can limit the defender's ability to reconstruct or reacquire the initial weights that are perfectly aligned with the modified model's architecture. In this paper, for the sake of comparison with methods that require pre-trained weights, we will take BERT\cite{BERT} as an example.
\subsection{MBTSAD}
\begin{algorithm}{}
    \caption{MBTSAD}
    \label{MBTSAD_alg}
    \KwIn{Clean dataset $D_c = \{(x_1,y_1), \ldots ,(x_n,y_n)\}$\, the parameters 
$\theta_b$ for backdoored model $M_b$, The number of augmented texts per sample $N$.}
    \KwOut{backdoor-free model's parameters $\theta_c$.}
    \tcp{Generate Augmented Dataset $D_a$}
        $D_a \leftarrow \emptyset$ \tcp*{Initialize $D_a$}
        \ForEach {$(x_i, y_i)$ in $D_c$}{
            \For{$k=1$ \KwTo $N$}{
                $x'_{i,k} \leftarrow \text{Token Splitting}(x_i)$\;
                $D_a \leftarrow D_a \cup \{(x'_{i,k}, y_i)\}$\; 
                \tcp{Add augmented samples to $D_a$}
            }
        }
    \tcp{Retrain $M_b$ on $D_a$}
    $\theta_p \gets \theta_b$ \;
    \While{no convergence}{
        Compute the gradient $\nabla_{\theta_p }\mathcal{L}_{ce}(\theta_p, y)$ \;
        $\theta_p \gets \theta_p - \delta \nabla_{\theta_p} \mathcal{L}_{ce}(\theta_p, y)$\;
    }
    \tcp{Attention Distillation}
    \While{no convergence}{
        Compute $\nabla _{\theta_b }\mathcal{L}_{total}(\theta_p, \theta_b, y)$ 
 and $\mathcal{L}_{total}$ in \eqref{total loss}\;
        $\theta_b \gets \theta_b - \gamma \nabla_{\theta_b} \mathcal{L}_{total}(\theta_p, \theta_b, y)$\;
    }
    $\theta_c \gets \theta_b$ \;
        \Return $\theta_c$
\end{algorithm}
Fig. \ref{MBTSAD-picture}  and Algorithm \ref{MBTSAD_alg} illustrate the process of MBTSAD. The key steps of the proposed MBTSAD include: 1) retraining the backdoor model \(M_b\) on an augmented dataset \(D_a\) generated by token splitting to obtain a model \(M_p\) that preliminarily mitigates the backdoors. $\delta$ is the learning rate. The parameters of $M_p$ and $M_b$ are denoted as $\theta_p$ and $\theta_b$, respectively. Notably, for each text \( x_i \) in \( D_c \), the token splitting algorithm is employed to create \( N \) different augmented texts, all labeled identically to \( x_i \). If the number of clean data is \( |D_c| \), then the number of samples in the augmented dataset \( D_a \) is \( N|D_c| \). 2) employing attention distillation to mitigate backdoors further and get a backdoor-free model \(M_c\). In the first step, the augmented dataset \(D_a\) is generated from a small subset dataset \(D_c\) of the full clean dataset. In the second step, the backdoored model \(M_b\) acts as the student model, while the model \(M_p\) serves as the teacher model. The learning rate is $\gamma$ and the dataset used in the attention distillation is the augmented dataset \(D_a\). MBTSAD distills various components of the model, including the embedding layer, attention layer, and classification head. Next, we will provide a detailed explanation of the token splitting and attention distillation within MBTSAD.
\subsubsection{Token Splitting}
For a text \( S = t_1, t_2, t_3, \ldots, t_n \) in the clean dataset, where \( S \) consists of \( n \) tokens. \( m \) is the number of target tokens. Each target token is split into two tokens in its original position to generate \(S'\). For a target token \( t_i=c_1 c_2 c_3 \ldots c_l \) with \(l\) characters, \(t_i\) split from a random position \(p\) into \(t'_1 = c_1 c_2 c_3 \ldots c_p\) and \(t'_2 = c_{p+1} \ldots c_l\). For example, consider the sentence “a stirring, funny and finally transporting re-imagining of beauty” in Fig.~\ref{MBTSAD-picture}. After applying the token splitting, it transforms into “a st irring, fu nny and finally transporting re-im gaining of be auty”. 

\subsubsection{Attention Distillation}
Retraining the backdoored model $M_b$ with dataset $D_a$ has resulted in a model $M_p$, where the backdoor attack success rate has been preliminarily reduced. Fig. \ref{MBTSAD-picture} shows that attention distillation involves using  $M_b$ as the student model and $M_p$ as the teacher model to obtain the backdoor-free model $M_c$. Attention distillation can further eliminate the backdoor patterns in the backdoored model while maintaining performances on clean data.

Related works found that attention weights learned by BERT can capture rich linguistic knowledge\cite{rich-info-transformer} \cite{TinyBERT}. This type of linguistic expertise encompasses syntactic details that are crucial for comprehending natural language.
Motivated by previous work about distillation\cite{TinyBERT},  the distillation loss is composed of attention transfer, hidden-states-based distillation, and classification loss, which is defined as
\begin{equation}
\mathcal{L}_{total} = \mathcal{L}_{a} + \alpha \cdot \mathcal{L}_{cls} + \beta \cdot \mathcal{L}_{h},
\label{total loss}
\end{equation}
where $\mathcal{L}_{total}$ denotes the loss of attention distillation, $\mathcal{L}_a$ denotes the loss of attention transfer, $\mathcal{L}_h$ denotes the loss of the hidden states, $\mathcal{L}_{cls}$ denotes the classification loss. Hyperparameters $\alpha, \beta$ are weights for loss functions.

MBTSAD employs attention-based attention transfer\cite{Attetion-based-attention-transfer}. Attention-based attention transfer facilitates a knowledge distillation process that is more attuned to regions of high-intensity activation. Related work has shown the effectiveness of attention-based attention transfer in defending against image backdoor attacks\cite{NAD}. It is observed that the attention mechanism in the backdoored language model effectively learns to associate triggers with target labels. Consequently, it is theoretically plausible to mitigate the backdoor patterns within the attention-base model through the application of attention-based attention transfer. In MBTSAD, the loss of attention-based attention transfer is formulated as

\begin{equation}
\mathcal{L}_a = \sum\limits_{n=1}^{N_L} AT(a^{n}_{t}, a^{n}_{s}) \label{at loss},
\end{equation}
where $a^{n}_{t} \in R^{c \times l \times l}, a^{n}_{s} \in R^{c \times l \times l}$ are the attention matrix corresponding to the $n-th$ transformer layer of teacher model $M_p$ and student model $M_b$ respectively. $c$ denotes the number of attention heads in each layer. $l$ is the input text length. $AT(\cdot)$ means the attention-based attention transfer loss function. We define Attention Enhancement (AE) function\cite{NAD} $AE(\cdot): R^{c \times l \times l} \rightarrow{R^{l \times l}}$. Attention Enhancement is capable of accentuating the elevated activation values within the attention matrix, which facilitates the capture of inter-word correlations. $AE(\cdot)$ is formulated as

\begin{equation}
AE(a^n) = \sum\limits_{j=1}^{c} |(a^{n, j})|^2 \label{ae func},
\end{equation}

where $a^n$ is the attention matrix corresponding to the $n-th$ transformer layer. $a^{n, j}$ is the $j-th$ head in $n-th$ transformer layer. $|\cdot|$ is the absolute value function. In this work, the softmax outputs of attention matrices are used instead of the unnormalized attention matrices. Therefore, $AT(\cdot)$ is formulated as

\begin{equation}
    AT(a^{n}_{t}, a^{n}_{s})= \left\|\frac{AE(a^{n}_{t})}{\left\|AE(a^{n}_{t})\right\|_{2}}-\frac{AE(a^{n}_{s})}{\left\|AE(a^{n}_{s})\right\|_{2}}\right\|_{2}, \label{at func}
\end{equation}
where $\left\|\cdot\right\|_{2}$ is the $L_2$ norm.

Both classification loss and hidden-states-based loss are introduced in MBTSAD to maintain the performance on clean samples. Hidden states encompass the outputs of all transformer layers and the word embedding layer. 
\begin{equation}
h^n=\left\{
\begin{array}{rcl}
h_{emb}       &      & {n = 0},\\
h^{n} _{trans}     &      & {0 < n \leq N_L},\\
\end{array} \right. \label{hidden states}
\end{equation}
where $h^n$ denotes the hidden states, $h_{emb}$ denotes the state of word embedding layer, $h^{n}_{trans}$ denotes the state of the $n-th$ transformer layer, $N_L$ denotes the number of transformer layer. Therefore, the loss of hidden states is formulated as

\begin{equation}
\mathcal{L}_h = \sum\limits_{n=0}^{N_L} MSE(h^{n}_{t}, h^{n}_{s}) \label{hidden states loss},
\end{equation}
where $h^{n}_{t}, h^{n}_{s}$ are the $n-th$ hidden state of teacher model $M_p$ and student model $M_b$, respectively. $MSE(\cdot)$ is the mean squared error loss function.
Classification loss is defined as 

\begin{equation}
\mathcal{L}_{cls} =  CE(Y_{out}, Y) \label{cls loss},
\end{equation}
where $CE(\cdot)$ is the cross entropy loss function, $Y_{out}$ is the predictions and $Y$ is the ground-truth.
\subsection{Theoretical Analysis of Token Splitting}\label{ts-section}
The first step in the MBTSAD involves the process of retraining using an augmented dataset $D_a$. This step can preliminarily mitigate backdoors, and adopting the TS methods in this step yields the best backdoor mitigation results. This phenomenon can be explained by adversarial training theory.

We modify the optimization objective function in \cite{understand-adv-learning} for different augmentation methods and it is defined as 
the following min-max formulation:
\begin{equation}
\begin{aligned}
& \min \limits_{\theta} \mathbb{E}_{(x, y) \sim \mathcal{D}}\left[\max \limits_{Aug \in B} \mathcal{L}_{ce}(\theta, Aug(x), y)\right]\\
& \text{s.t.} \quad \forall (x, y) \sim \mathcal{D}, \quad Q(Aug(x), x) < \delta, \\
\end{aligned}
\label{eqn:adv_training}
\end{equation}
where $(x, y) \sim \mathcal{D} $ represents training data sampled from distribution $\mathcal{D}$ , $Aug$ is the augmentation, $Q$ is the quantization function of perturbation and $B$ is the allowed augmentation methods set. However, if we directly solve the original problem, where each sample can be augmented using one of the available methods, the search space becomes \(O(|B|^{|D_c|})\), which is extremely large. To simplify this issue, we restrict the dataset to applying only one single augmentation method, applied only once before retraining. Therefore, as shown in \eqref{simple-adv-training}, adversarial training is simplified into two steps: 1) select the augmentation method \( Aug \) that maximizes the perturbation in the backdoored model's predictions, and 2) use the dataset \( D_a \) generated by \( Aug \) for retraining.
\begin{equation}
\label{simple-adv-training}
\begin{aligned}
& \min \limits_{\theta} \mathbb{E}_{(x, y) \sim \mathcal{D}}\left[\mathcal{L}_{ce}(\theta, Aug(x), y)\right] \\
&\text{s.t.} \quad \begin{cases}
    {Aug} = \arg \max \limits_{F \in B} \mathbb{E}_{(x, y) \sim \mathcal{D}}  \left[\mathcal{L}_{ce}(\theta_b, F(x), y)\right]\\
    \forall (x, y) \sim \mathcal{D} \;and \; \forall F \in B, Q(F(x), x) < \delta 
    \end{cases}
\end{aligned}
\end{equation}
where $\theta_b$ is the parameters of $M_b$. This simplification reduces the search space to \( O(|B|) \).
Compared with other data augmentation methods\cite{eda}\cite{karimi-etal-2021-aeda-easier}\cite{cwea}, the TS method meets the constraint condition that the backdoored model's prediction results are maximally perturbed. This perturbation will enable the model to learn more generalized features, thereby achieving the best backdoor mitigation results. In section \ref{exp-analysis-ts}, we will verify the above assumption and provide a more detailed analysis through the visualization of text representations and the analysis of cross-entropy loss.
\section{Experiments}
\subsection{Experimental Setup}
\subsubsection{Backdoor Attacks}
We adopt the BadNets\cite{BadNets}, LWP\cite{LWP}, and EP \cite{EP} to attack the BERT\cite{BERT} model. LWP poisons the lower layer weight. EP hacks the model in a data-free way by modifying one single word embedding vector, with almost no accuracy sacrificed on clean samples\cite{EP}. 

\subsubsection{Data Knowledge} We use the two-class Stanford Sentiment Treebank (SST-2) dataset\cite{SST-2}, the IMDb movie reviews dataset\cite{IMDb}. The defender has access to 20\% (\(r = 0.2\)) of the entire clean training dataset. 

\subsubsection{Baseline Defense Methods} To evaluate the effectiveness of MBTSAD, we adopt the state-of-art defense methods, including RAP\cite{RAP}, STRIP\cite{STRIP}, clean data fine-tuning (CFT), ``reinitialization'', ``Fine-mixing''\cite{Fine-mixing}, and ``Fine-purifying''\cite{fine-prifying}. RAP and STRIP are the defense methods based on detecting backdoor samples and we follow the implementation and settings in Openbackdoor\cite{openbackdoor}. CFT fine-tunes the backdoored model $M_b$ on $D_c$. The ``reinitialization'' method reinitializes the weights in the last transformer layer of BERT.
The training dataset utilized in clean data fine-tuning, ``reinitialization'', ``Fine-mixing'', and ``Fine-purifying'' is the accessible clean training dataset $D_c$. Additionally, the results of retraining $M_b$ with $D_a$ (denoted as TS-FT) are also reported. 

\subsubsection{Metrics}
We adopt the accuracy (CACC /\%) on the clean test set and the backdoor attack success rate (ASR /\%) on the poisoned test set to measure the clean and backdoor performance. For each table of results, the best backdoor mitigation results with the lowest ASRs and the best results on clean data with the highest CACCs are marked in \textbf{bold}.
\subsubsection{Parameter Setup}
For backdoor attacks, the backdoored language model \(M_b\) is trained on a poisoned dataset with a poisoning rate of 0.1. We use the number of target tokens to perturb each text as the quantization function of perturbation $Q$. For the TS method, we follow the settings in \cite{ma2019nlpaug}, and for a text with \(n\) tokens, the number of target tokens is $max(10, 0.3n)$. Each clean sample in $D_c$ can generate \(N = 3\) different augmented samples. The learning rate in retraining is $2 \times 10^{-5}$ and in attention distillation is $5 \times 10^{-4}$. For $\alpha$ and $\beta$ in \eqref{total loss}, we adaptively set it to different values for each backdoor attack.

\subsection{Main Results}
\subsubsection{Detection-Based Method}
\begin{table}[htbp]
\caption{Results of Different Detection-based Defense Methods}
\begin{center}		
  \begin{tabular}{|c|c|c|c|c|c|}
  \hline
			\multicolumn{1}{|c|}{\multirow{2}{*}{\textbf{Attack}}}	&\multicolumn{1}{c|}{\multirow{2}{*}{\textbf{Defense}}}	& \multicolumn{2}{c|}{\textbf{SST-2}}     & \multicolumn{2}{c|}{\textbf{IMDb}}\\
    \cline{3-6}
    & & \textit{CACC} & \textit{ASR} & \textit{CACC} & \textit{ASR}\\
    \hline
\multirow[m]{4}{*}{BadNets}	& None$^{\mathrm{a}}$	& 91.32	& 89.88 & 93.48 & 89.94\\
			  	    & STRIP	& \textbf{91.49}	& 39.36 & \textbf{93.34} &70.94 \\
			             & RAP	& 90.66	& 32.35 & 91.24 & 48.36\\
                   & \textbf{MBTSAD}			&90.17 	&\textbf{9.65}  & 92.92  & \textbf{17.24} \\
                   \hline
\multirow[m]{4}{*}{LWP}	& None	& 90.66	& 94.41 & 92.87 & 95.50\\

			  	    & STRIP	& \textbf{90.88}	& 93.73 & 89.13  & 91.12\\
			             & RAP	& 90.77	& 53.27 & 91.73 & 62.52\\
                & \textbf{MBTSAD}			& 90.72			& \textbf{15.90}  & \textbf{93.01} & \textbf{16.26} \\
                   \hline
\multirow[m]{4}{*}{EP}	& None	& 91.21	& 95.17 & 94.03 & 98.66\\
			  	    & STRIP	& 87.69	& 93.20 & \textbf{93.85} & 98.59\\
			             & RAP	& 87.86	& 93.41 & 71.81 & 51.45\\
    & \textbf{MBTSAD}			& \textbf{89.07}			&\textbf{12.83} & 93.65 & \textbf{9.64}  \\
    \hline
\multicolumn{4}{l}{$^{\mathrm{a}}$No defense.}
		\end{tabular}
\label{backdoor detection table}
\end{center}
\end{table}
We compare MBTSAD with RAP and STRIP. The experimental results are presented in Table \ref{backdoor detection table}.
STRIP can not defend LWP and EP attacks effectively and the ASR on the SST-2 dataset is 93.73\% and 93.20\%. RAP falls short in countering EP attacks and significantly degrades the model's clean performance. For instance, on the SST-2 dataset, the CACC drops to 87.86\%. In contrast, MBTSAD achieves the best defense performance on Badnets, LWP, and EP attacks, reducing the ASR to below 20\% and even 9.65\% in the SST-2 dataset.

\subsubsection{Backdoor Mitigation Methods}
\begin{table}[htbp]
\caption{Results of Different Backdoor Mitigation Methods}
\begin{center}		
  \begin{tabular}{|c|c|c|c|c|c|}
  \hline
			\multicolumn{1}{|c|}{\multirow{2}{*}{\textbf{Attack}}}	&\multicolumn{1}{c|}{\multirow{2}{*}{\textbf{Defense}}}	& \multicolumn{2}{c|}{\textbf{SST-2}}     & \multicolumn{2}{c|}{\textbf{IMDb}}\\
    \cline{3-6}
    & & \textit{CACC} & \textit{ASR} & \textit{CACC} & \textit{ASR}\\
    \hline
\multirow[m]{8}{*}{BadNets}	& None$^{\mathrm{a}}$	& 91.32	& 89.88 & 93.48 & 89.94\\
                & CFT $^{\mathrm{c}}$			& \textbf{91.82}	& 65.02  & \textbf{93.18}  & 85.77 \\ 

                & Reinitialization			& 89.95			& 27.96 & 93.19 & 37.29 \\
                & Fine-Pruning		&  89.29			& 16.18 & 92.35 & 52.11 \\
                & TS-FT	$^{\mathrm{d}}$		& 90.45		&33.38  &93.20  &38.46 \\
                & \textbf{MBTSAD}			&90.17 	&\textbf{9.65}  & 92.92  & \textbf{17.24} \\
                \rowcolor{gray!30}& Fine-Mixing$^{\mathrm{b}}$			& 88.34			& 15.79 & 92.23 & 9.10\\ 
                \rowcolor{gray!30}& Fine-Purifying$^{\mathrm{b}}$			& 88.74			& 16.01 &91.93  &9.69 \\ 
                   \hline
\multirow[m]{8}{*}{LWP}	& None	& 90.66	& 94.41 & 92.87 & 95.50\\
                & CFT			& \textbf{91.43}	& 93.43  &91.86  & 95.47\\
                & Reinitialization			& 89.46			& 47.15 &92.19  &92.24 \\
                & Fine-Pruning		& 88.41			& 18.53 &91.47  &27.14 \\
                 & TS-FT			&90.74 			&25.98  & 92.88    & 30.80 \\
                & \textbf{MBTSAD}			& 90.72			& \textbf{15.90}  & \textbf{93.01} & \textbf{16.26} \\
                \rowcolor{gray!30}& Fine-Mixing			& 88.91			&12.39  &92.37  &7.56 \\
                \rowcolor{gray!30}& Fine-Purifying		& 88.57			& 15.13 &92.36  &8.21 \\
                   \hline
\multirow[m]{8}{*}{EP}	& None	& 91.21	& 95.17 & 94.03 & 98.66\\
                & CFT			& \textbf{91.56}			& 93.78 & 93.70  & 84.03 \\ 
                & Reinitialization			& 89.51			&90.02 &93.69  &82.75 \\
                & Fine-Pruning		& 87.59			& 78.40 &92.96  &37.95 \\
                & TS-FT			&90.88 			&38.38  &93.65  &40.15 \\
                & \textbf{MBTSAD}			& 89.07			&\textbf{12.83} & 93.65 &\textbf{ 9.64}  \\
                \rowcolor{gray!30}& Fine-Mixing			&88.52
                &11.18  &92.33  &9.30 \\
                \rowcolor{gray!30}& Fine-Purifying		& 88.19		& 14.39 &92.10  &7.75 \\            
    \hline
\multicolumn{6}{l}{$^{\mathrm{a}}$No defense, $^{\mathrm{b}}$Methods based on pre-trained weights}\\
\multicolumn{6}{l}{$^{\mathrm{c}}$Fine-tuning on clean data, $^{\mathrm{d}}$Retraining on augmented dataset $D_a$}
\end{tabular}
\label{backdoor_mitigation_table}
\end{center}
\end{table}
The results of MBTSAD and other backdoor mitigation methods are presented in Table \ref{backdoor_mitigation_table}. It is evident that direct fine-tuning with clean data $D_c$ can mitigate Badnets attacks with an ASR drop of nearly 25\% on the SST-2 dataset, but it struggles to defend against LWP and EP attacks effectively. The ``reinitialization'' method presumes that fine-tuning with poisoned data primarily affects the high-level layer weights, rendering it ineffective against low-level layer attacks like LWP and EP that target word embeddings. For instance, in the LWP attacks on the IMDb dataset, the ASR remains as high as 92.24\%. Fine-pruning can yield promising performance on mitigating backdoors, but the limited amount of available clean data (20\%) significantly degrades the CACC. 

In contrast, our MBTSAD method demonstrates the ability to defend against all three attacks. Its efficacy in backdoor mitigation is comparable to methods based on pre-trained weights. Notably, on the SST-2 dataset, MBTSAD outperforms ``Fine-mixing'' and ``Fine-purifying'' in defending against Badnets attacks. Furthermore, compared to retraining on a dataset augmented by the TS (the first step in MBTSAD, denoted as ``TS-FT'' in Table \ref{backdoor_mitigation_table}), MBTSAD achieves superior backdoor mitigation performance without a considerable compromise in CACC. This highlights the effectiveness of the attention distillation proposed in MBTSAD.
\subsection{Ablation Study}
\subsubsection{Experimental Analysis of Token Splitting}\label{exp-analysis-ts}
We evaluate the performance of backdoor mitigation for retraining when using different augmented methods (the first step in MBTSAD). The data augmentation methods are EDA\cite{eda}, AEDA\cite{karimi-etal-2021-aeda-easier}, CWEA\cite{cwea}, and ``Add Trig'', which means $B =$ \{TS, AEDA, EDA, Add Trig, CWEA\} in \eqref{simple-adv-training}. ``Add Trig'' refers to randomly inserting  ``cb'' into the text. ``cb'' is not in the LWP trigger set, thus simulating a scenario where the defender is unaware of the backdoor trigger in advance. Table \ref{table-diff-aug-result} indicates that the TS method provides the best defense against all three attacks. 

\begin{figure*}[htbp]
\centering
    \includegraphics[scale=.32]{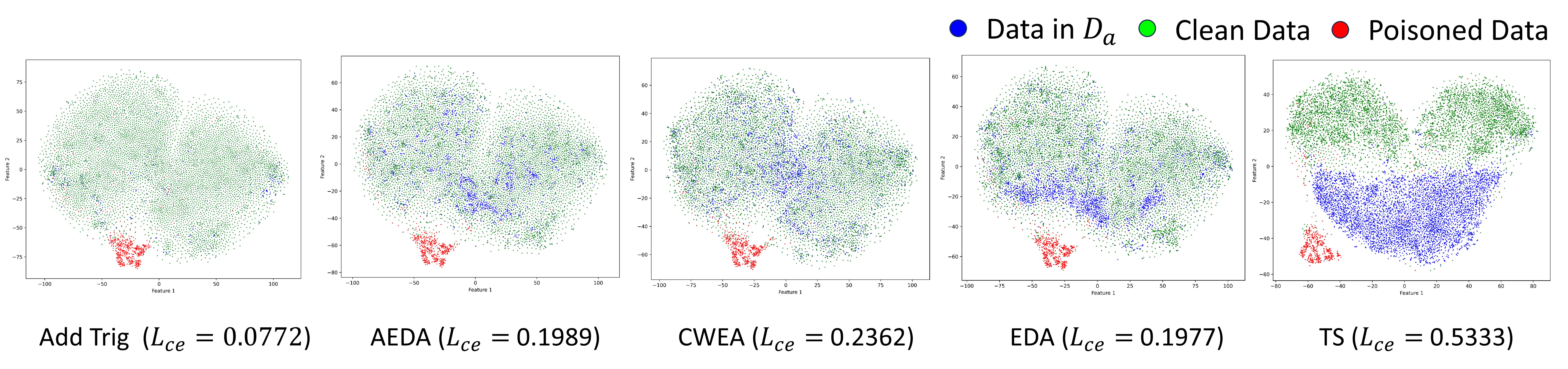}
\caption{The results of text representations visualization for datasets from different data augmentation methods including EDA, AEDA, CWEA, Add Trig, and token splitting under the settings of LWP attack. We calculate the average cross-entropy loss ($L_{ce}$) for each batch in SST-2 dataset using the LWP backdoored model, with a batch size of 16, and the loss is shown in parentheses. The blue represents the data augmented dataset $D_a$. The green and red represent the clean data and poisoned data in $D_{poison}$, respectively.}
\label{embs-picture}
\end{figure*}

Then, we poison the SST-2 dataset to obtain the backdoored dataset \( D_{poison} \). The augmented dataset \( D_{a} \) is generated based on all clean data from \( D_{poison} \) and each text generates one augmented text. In Fig. \ref{embs-picture}, the TS method results in the highest loss, averaging 0.5333. According to the simplified adversarial training defined in  \eqref{simple-adv-training}, compared to other augmentation methods, TS introduces the most significant perturbations to the original clean dataset and consequently gets the best backdoor mitigation results.

We further visualize the text representations and analyze the semantic distribution of the dataset. Specifically, we use the backdoored model for inference and visualize the [CLS] token outputs using t-SNE\cite{t-sne}. As shown in Fig. \ref{embs-picture} indicates the [CLS] vector distributions of data generated by methods other than TS are similar to those of clean samples from $D_{poison}$. The perturbation generated by TS causes the distribution of these samples in the semantic space to be inconsistent with that of the clean training data, rendering them as OOD data\cite{OOD-nlp}. Moreover, the labels remain clean. Consequently, these OOD data enable the model to learn more generalized features, which helps mitigate the strong association between the backdoor triggers and the target label, thereby reducing the impact of the backdoor attack.

\begin{table}
\caption{Retraining Results of Different Augmentation Methods on SST-2}
\label{table-diff-aug-result}
\begin{center}		
  \begin{tabular}{|c|c|c|c|}
  \hline
			\textbf{Attacks}	& \textbf{Methods}	& \textbf{CACC} & \textbf{ASR}\\
			\hline
\multirow[m]{6}{*}{BadNets}	& None$^{\mathrm{a}}$			& 91.32			& 89.88\\
			  	        & EDA 			& 90.17			& 27.19\\
			                 & AEDA			& \textbf{90.77}			& 42.54\\
                    & CWEA			&87.53 			& 94.85\\
			  	        & Add Trig $^{\mathrm{b}}$		& 90.61		& 39.80\\
			                 & TS			& 90.45			& \textbf{33.38}\\
                   \hline
\multirow[m]{6}{*}{LWP}	& None			& 90.66			& 94.41\\
			  	& EDA 			& 90.56			& 87.28\\
			                 & AEDA			& 91.38		& 81.36\\
                    & CWEA			& 89.62			& 88.82\\
			  	        & Add Trig 			& 90.66			& 93.75\\
			                 & TS 			& \textbf{90.74}			& \textbf{25.98}\\
                   \hline
\multirow[m]{6}{*}{EP}    & None			& 91.21			& 95.17\\
			  	    & EDA			& 89.78			& 93.86\\
			                 & AEDA			& 89.84			& 81.36\\
                    & CWEA			& 88.69			& 69.30\\
			  	        & Add Trig			& 90.28			& 93.53\\
			                 & TS			& \textbf{90.88}		& \textbf{38.38}\\
			\hline
\multicolumn{4}{l}{$^{\mathrm{a}}$No defense, $^{\mathrm{b}}$Randomly add ``cb'' in text.}
\end{tabular}
\end{center}
\end{table}

\subsubsection{Loss Function} We explore the relation among the attention transfer loss term $\mathcal{L}_a$, hidden states loss term $\mathcal{L}_h$, and classification loss term $\mathcal{L}_{cls}$ within the distillation loss $\mathcal{L}_{total}$. We designed three ablation methods to compare the effects: one without the $\mathcal{L}_a$ (denoted as MBTSAD-no-att), one without the $\mathcal{L}_h$ (denoted as MBTSAD-no-hid), and the distillation with only $\mathcal{L}_{cls}$ (denoted as CD). All other hyperparameters are kept consistent with MBTSAD. The results, as presented in Table \ref{ablation-result}, indicate that MBTSAD outperforms in almost all scenarios. Furthermore, CD does not achieve promising results by relying solely on the $\mathcal{L}_{cls}$ to mitigate backdoors. For instance, on the SST-2, the ASR is only reduced to 53.83\%, 47.26\%, and 50.33\% respectively. Lastly, Integrating $\mathcal{L}_{cls}$ and only $\mathcal{L}_{a}$ (MBTSAD-no-hid) can effectively mitigate backdoors but results in a significant decrease in clean performance, with a 5.74\% drop in CACC on the IMDb dataset under the LWP attack. On the other hand, integrating $\mathcal{L}_{cls}$ and only $\mathcal{L}_{h}$ (MBTSAD-no-att) performs the best in CACC, but results in a significant increase in ASR, with a 24.01\% increase in ASR compared with MBTSAD on the SST-2 dataset under the EP attack.
MBTSAD balances the drop of CACC and the mitigation of backdoors, for instance, reducing the ASR to 9.64\% with only a 0.38\% decline in CACC compared with the original backdoored model on the IMDb dataset under the EP attack. In conclusion, the integration of the attention loss term, hidden states loss term, and classification loss term are essential for effectively mitigating backdoors while minimizing the decline in clean performance.
\begin{table}
\caption{Results of Ablation Defense Methods}
\label{ablation-result}
\begin{center}		
  \begin{tabular}{|c|c|c|c|c|c|}
  \hline
			\multicolumn{1}{|c|}{\multirow{2}{*}{\textbf{Attack}}}	&\multicolumn{1}{c|}{\multirow{2}{*}{\textbf{Defense}}}	& \multicolumn{2}{c|}{\textbf{SST-2}}     & \multicolumn{2}{c|}{\textbf{IMDb}}\\
    \cline{3-6}
    & & \textit{CACC} & \textit{ASR} & \textit{CACC} & \textit{ASR}\\
    \hline
\multirow[m]{5}{*}{BadNets}	& None$^{\mathrm{a}}$	& 91.32	& 89.88 & 93.48 & 89.94 \\
                            & \textbf{MBTSAD}	& 90.17	&\textbf{9.65}  & 92.92  & \textbf{17.24} \\
			  	    & MBTSAD-no-hid	& 85.89	& 32.68 &\textbf{93.60}  &25.86  \\
			             & MBTSAD-no-att	&88.96 	& 17.43 & 92.26  &15.66  \\
                        & CD $^{\mathrm{b}}$	&\textbf{92.03} 	&53.83  &93.46 &36.15 \\
                   \hline
\multirow[m]{5}{*}{LWP}	& None	& 90.66 	&94.41  & 92.87  & 95.50 \\
                            & \textbf{MBTSAD}	&\textbf{90.72}  	& \textbf{15.90}  & 93.01  & 16.26 \\
			  	    & MBTSAD-no-hid	& 90.66	&25.11  &87.13  &\textbf{12.15}  \\
			             & MBTSAD-no-att	&88.85 	&16.67  &\textbf{93.04}  &37.69  \\
                        & CD	& 88.91 	&47.26  &91.85 & 63.75 \\
                   \hline
\multirow[m]{5}{*}{EP}	& None	& 91.21	& 95.17  & 94.03  & 98.66 \\ 
                            & \textbf{MBTSAD}	&89.07  	&\textbf{12.83}   & 93.65  & \textbf{9.64} \\
			  	    & MBTSAD-no-hid	& 87.59  &45.72  &93.79  & 39.06\\
			             & MBTSAD-no-att & \textbf{90.77}	& 36.84 & \textbf{94.11} &20.30 \\
                        & CD	&89.18 	&50.33  &93.40  &18.97 \\
    \hline
\multicolumn{6}{l}{$^{\mathrm{a}}$No defense, $^{\mathrm{b}}$the distillation with only $\mathcal{L}_{cls}$}		
\end{tabular}
\end{center}
\end{table}

\section{Future Work}
We primarily investigate word-level backdoor attacks (Badnets, LWP, and EP) in our experiments, without delving into backdoor attacks at other levels. Additionally, while MBTSAD is effective for models with attention mechanisms, its efficacy may not extend to models with different architectures. Our future work will focus on two areas: exploring MBTSAD's potential to defend against sentence-level backdoor attacks and designing a more general backdoor defense method independent of the attention mechanism.
\section{Conclusion}
In this paper, we propose MBTSAD, a method for mitigating backdoors in language models using only 20\% clean data and no pre-trained weights. MBTSAD involves two steps: 1) retraining the backdoored model on a dataset generated by token splitting; 2) using the retrained model as a teacher and the backdoored model as a student, and applying attention distillation to obtain a backdoor-free model. Experimental results on SST-2 and IMDb datasets show MBTSAD achieves superior backdoor mitigation with minimal clean performance loss. Ablation studies confirm the necessity of the attention loss term, hidden states loss term, and classification loss term in attention distillation. 

Additionally, adversarial training theory and text representations visualization reveal that the TS method in MBTSAD's first step generates Out-Of-Distribution data, leading the model to learn more generalized features and eliminate backdoor patterns. We hope that MBTSAD can provide a robust baseline method for defending against backdoor attacks in language models.

\bibliographystyle{IEEEtran}
\bibliography{references}
\end{document}